\documentclass[%
 aip,
 amsmath,amssymb,
 reprint,%
]{revtex4-1}

\usepackage{graphicx}
\usepackage{dcolumn}
\usepackage{bm}

\usepackage[utf8]{inputenc}
\usepackage[T1]{fontenc}
\usepackage{etoolbox}
\usepackage{hyperref}
\usepackage{rotating}
\usepackage{xcolor}
\usepackage{xr}
\usepackage{wrapfig}
\usepackage{amsmath}
\usepackage{enumitem}
\usepackage{notoccite}
\usepackage{comment}

\newcommand*{\hatH}{\hat{\mathcal{H}}}

\makeatletter
\def\@email#1#2{%
 \endgroup
 \patchcmd{\titleblock@produce}
  {\frontmatter@RRAPformat}
  {\frontmatter@RRAPformat{\produce@RRAP{*#1\href{mailto:#2}{#2}}}\frontmatter@RRAPformat}
  {}{}
}%
\makeatother

\begin{document}

\preprint{AIP/123-QED}

\title{Spin-phonon interactions and magnetoelectric coupling 
in Co$_4$$B_2$O$_9$ ($B$ = Nb, Ta)}

\author{K. Park}
\affiliation{Department of Chemistry, University of Tennessee, Knoxville, Tennessee 37996, USA}

\author{J. Kim}
\affiliation{Department of Physics and Astronomy, Rutgers University, Piscataway, New Jersey 08854, USA}

\author{S. Choi}
\affiliation{Department of Physics and Astronomy, Rutgers University, Piscataway, New Jersey 08854, USA}
\affiliation{Center for Integrated Nanostructure Physics, Institute for Basic Science, Suwon 16419, Republic of Korea}
\affiliation{Sungkyunkwan University, Suwon 16419, Republic of Korea}

\author{S. Fan}
\affiliation{Department of Chemistry, University of Tennessee, Knoxville, Tennessee 37996, USA}

\author{C. Kim}
\affiliation{Department of Energy and Chemical Engineering, Ulsan National Institute of Science and Technology (UNIST), Ulsan, Korea}

\author{D. G. Oh}
\affiliation{Department of Physics, Yonsei University, Seoul 03722, Korea}

\author{N. Lee}
\affiliation{Department of Physics, Yonsei University, Seoul 03722, Korea}

\author{S. -W. Cheong}
\affiliation{Department of Physics and Astronomy, Rutgers University, Piscataway, New Jersey 08854, USA}
\affiliation{Rutgers Center for Emergent Materials, Rutgers University, Piscataway, New Jersey 08854 USA}

\author{V. Kiryukhin}
\affiliation{Department of Physics and Astronomy, Rutgers University, Piscataway, New Jersey 08854, USA}

\author{Y. J. Choi}
\affiliation{Department of Physics, Yonsei University, Seoul 03722, Korea}

\author{D. Vanderbilt}
\affiliation{Department of Physics and Astronomy, Rutgers University, Piscataway, New Jersey 08854, USA}

\author{J. H. Lee}
\affiliation{Department of Energy and Chemical Engineering, Ulsan National Institute of Science and Technology (UNIST), Ulsan, Korea}

\author{J. L. Musfeldt}
\email{musfeldt@utk.edu}
\affiliation{Department of Chemistry, University of Tennessee, Knoxville, Tennessee 37996, USA}
\affiliation{Department of Physics and Astronomy, University of Tennessee, Knoxville, Tennessee 37996, USA}

\date{\today}

\begin{abstract}

In order to explore the consequences of spin-orbit coupling on spin-phonon interactions in a set of chemically-similar mixed metal oxides, we measured the infrared vibrational properties of Co$_4B_2$O$_9$ ($B$ = Nb, Ta) as a function of temperature and compared our findings with 
lattice dynamics calculations and several different models of spin-phonon coupling. 
Frequency vs. temperature trends for the Co$^{2+}$ shearing mode near 150 cm$^{-1}$ reveal significant shifts across the magnetic ordering temperature that are especially large in relative terms.
Bringing these results together and accounting for noncollinearity, we obtain spin-phonon coupling constants of -3.4  and -4.3 cm$^{-1}$ for Co$_4$Nb$_2$O$_9$ and  the Ta analog, respectively. Analysis reveals that these coupling constants derive 
from interlayer (rather than intralayer) exchange interactions and that the interlayer interactions contain competing antiferromagnetic and ferromagnetic contributions. At the same time, beyond-Heisenberg terms are minimized due to fortuitous symmetry considerations, different than most other 4$d$- and 5$d$-containing oxides. Comparison with other contemporary oxides shows that spin-phonon coupling in this family of materials is among the strongest ever reported, suggesting an 
origin for magnetoelectric coupling.

\end{abstract}

\maketitle


Magnetic materials hosting both transition metal centers and heavy elements are contemporary platforms for the study of chemical bonding and novel properties. The strategy is that 3$d$ ions deliver localized orbitals, high spin, and strong electron correlation, whereas 4$d$ and 5$d$ centers contribute more diffuse orbitals, greater hybridization, a tendency toward dimerization, and spin-orbit coupling that competes on an equal footing with electron correlations.\cite{Taylor2017,Zwartsenberg2020,Kim2022,Varotto2022} 
This competition endows these materials with remarkable properties 
including ultra-hard magnetism,\cite{Singleton2016,Oneal2019} two-sublattice magnetism with frustration\cite{Bordacs2009,Kant2009,Wysocki2016} or independent 
ground states,\cite{Morrow2013,Yan2014} and mixing across broad energy scales.\cite{Datta2022}
%
One important consequence of spin-orbit interactions in these systems is spin-phonon coupling, conventionally described in terms of how the exchange interactions are modulated by particular displacement patterns.\cite{Martins2017,Kim2020,Oneal2019,Sushkov2005,Bordacs2009,Kant2009,Wysocki2016,Fennie2006}
In addition to revealing how materials communicate across different energy scales,  these interactions can drive multiferroicity.\cite{Lee2010,Mochizuki2011,Spaldin2019}

\begin{figure*}[thb]
\begin{minipage}{7in}
\includegraphics[width = 7in]{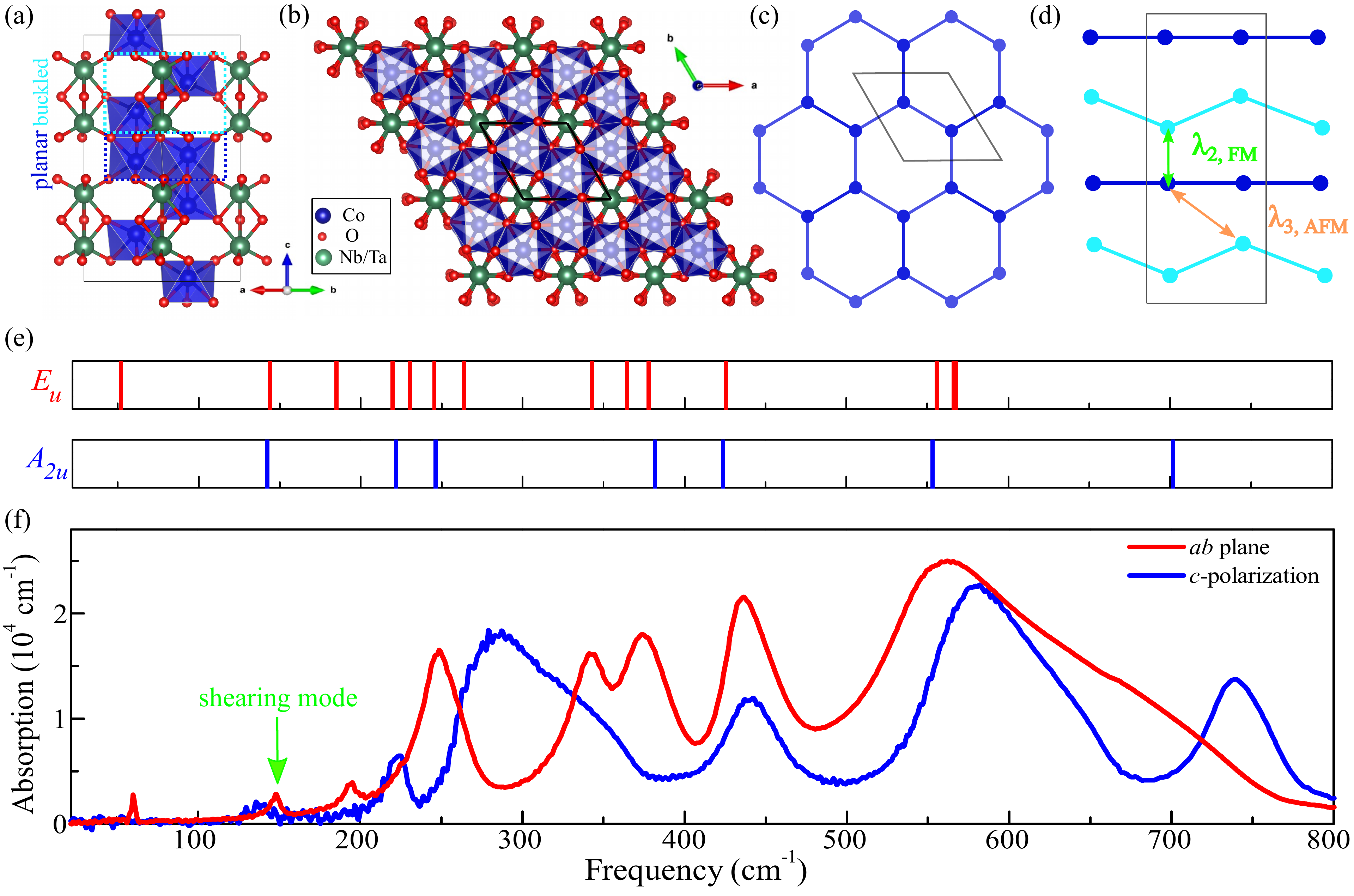}
\end{minipage}
\begin{minipage}{7in}
\caption{\label{Structure} (a, b) Crystal structure of Co$_4$Nb$_2$O$_9$.\cite{Cao2015,Lee2020,Choi2020} Blue, red, and green represent Co, O, and Nb/Ta centers. Octahedra are drawn only for magnetic ions. (c) A simplified view of the planar Co$^{2+}$ layer shows a honeycomb-like structure.  (d) Schematic illustrating  the planar and buckled layers. Only Co$^{2+}$ ions are shown. There are two relevant interlayer interactions  ($\lambda_{2,FM}$ and $\lambda_{3,AFM}$). 
(e)  Calculated $E_u$ and $A_{2u}$  phonon frequencies. (f) Polarized infrared absorption of Co$_4$Nb$_2$O$_9$  at 300 K. The response of Co$_4$Ta$_2$O$_9$ is similar and shown in Fig. S1 of the Supplementary information. 
}
\end{minipage}
\end{figure*}

The Co$_4B_2$O$_9$ ($B$ = Nb, Ta)  system is quasi-two dimensional mixed metal oxide with a $P\bar{3}c1$ space group [Fig. \ref{Structure}].\cite{Bertaut1961}
This corundum-type structure is derived from Cr$_2$O$_3$ such that four Cr sites are occupied by the magnetic Co$^{2+}$ ions and two nonmagnetic Nb$^{5+}$ or Ta$^{5+}$ ions reside on the $B$ sites. 
%
%
The octahedrally-coordinated Co$^{2+}$ centers are trigonally distorted and both edge- and face-sharing. The $B$ ions are arranged into vertical columns and occupy trigonally distorted octahedral sites in the buckled layer. 
Both systems order antiferromagnetically at 
$T_{\rm N}$ = 27 and 20 K for Co$_4$Nb$_2$O$_9$ and the Ta analog, 
respectively.\cite{Kolodiazhnyi2011,Cao2015} 
Originally thought to host collinear $ab$-plane spin structure  with a small moment along $c$,\cite{Bertaut1961,Khanh2016}  recent work establishes a noncollinear  spin arrangement  - with a different magnetic space group. 
\cite{Deng2018,Ding2020,Choi2020} 
A number of teams report magnetostriction across $T_{\rm N}$. 
\cite{Yin2016,Xie2016,Khanh2019} 
Under magnetic field, these materials reveal a spin-flop transition (at 0.3 T for $H$ $\parallel$ ${ab}$),\cite{Kolodiazhnyi2011,Cao2015,Yin2016} 
large magnetoelectric coupling,\cite{Fang2015,Fang2015a,Xie2016,Yin2016,Zhou2018} 
magnetodielectric behavior,\cite{Yadav2022,Zhou2018} a symmetry reduction with asymmetric distortion,\cite{Khanh2019} and spin excitations with magnetoelectric characteristics.\cite{Dagar2022}
The magnetoelectric coupling is approximately $\alpha_{[110]}$ = 20 ps/m for both compounds.\cite{Fang2015,Fang2015a,Xie2016,Yin2016} The interaction is nonlinear, \cite{Cao2017,Lee2020} 
and a polarization memory effect is  observed in the paramagnetic  phase.\cite{Xie2018} 
Despite evidence for magnetostriction across $T_{\rm N}$\cite{Xie2016,Khanh2016,Solovyev2016,Khanh2019} and proposals that cast spin-phonon coupling as the underlying mechanistic driver for these effects,\cite{Fang2015,Yin2016,Solovyev2016,Deng2018,Matsumoto2019} the fundamental excitations of the lattice and their changes across the magnetic ordering transitions are under-explored with only one Raman study of polycrystalline sample identifying modest coupling in two low-frequency modes.\cite{Yadav2022}
Co$_4$Nb$_2$O$_9$ and the Ta analog are also ideal platforms for unraveling structure-property relations - not just across simple trends but in higher level coupling processes as well.\cite{Chakarawet2021}

In order to explore these issues in greater depth, we combine polarized infrared reflectance, a symmetry analysis, and lattice dynamics calculations to reveal the vibrational properties of Co$_4$Nb$_2$O$_9$ and Co$_4$Ta$_2$O$_9$ across their magnetic ordering transitions. We find that the in-plane Co-containing shearing mode couples strongly to the spin system, red-shifting across $T_{\rm N}$ in both materials. The frequency shifts are extremely large, leading to coupling constants of -3.4 and -4.3 cm$^{-1}$ in Co$_4$Nb$_2$O$_9$ and the Ta analog, respectively. 
Remarkably, analysis of the spin-phonon coupling demonstrates that this displacement pattern modulates only the inter-plane magnetic interactions and that the latter contains competing antiferromagnetic and ferromagnetic terms. 
In addition to comparisons with  other contemporary oxides, we discuss how unique inter-layer spin-phonon interactions drive magnetoelectric coupling in this class of materials.


High-quality single crystals of Co$_4$Nb$_2$O$_9$ and Co$_4$Ta$_2$O$_9$  were grown by flux techniques.\cite{Lee2020,Choi2020} 
%
%
%
Near normal reflectance was measured over a wide frequency range (25 - 55,000 cm$^{-1}$) using a series of spectrometers including a Bruker 113v Fourier transform infrared spectrometer equipped with a liquid helium cooled bolometer detector, a Bruker Equinox 55 equipped with an infrared microscope, and a Perkin-Elmer Lambda-1050 grating spectrometer.  Appropriate polarizers revealed the $ab$-plane and $c$-axes response.  
A Kramers-Kronig analysis was used to obtain the optical constants.\cite{Wooten1972} 
We employed a constant low frequency extrapolation and a high frequency extrapolation of ${\omega}^{-1.75}$. 
The infrared absorption, ${\alpha}$($\omega$), and the real part of the optical conductivity, ${\sigma}_1$($\omega$), are of primary interest in this work.  
Open-flow cryostats provided temperature control. 
The theoretical phonon frequencies were calculated using the {\small{VASP}} code.\cite{Kresse96a,Kresse96b}
The pseudopotentials are of the projector-augmented-wave type as implemented in {\small{VASP}},\cite{Blochl94,KressePAW} 
with valence configurations $3d^7 4s^2$ for Co, $2s^2 sp^4$ for O, $4d^3 4p^6 5s^2$ for Nb, and $5d^3 6s^2$ for Ta.
The exchange-correlation functional is described by the Perdew-Burke-Ernzerhof type generalized gradient approximation,\cite{PBE}
with Dudarev type Hubbard $U$ correction\cite{Dudarev1998} on Co $3d$ orbits by 3 eV.
The plane-wave cut-off energy is set to 400 eV. The Brillouin zone sampling grid is $12 \times 12 \times 4$ including the $\Gamma$ point. Spin-orbit coupling is not taken into account. 
The structural coordinates are relaxed within a force threshold of 1.0 meV/\AA.
To obtain the $\Gamma$ point phonon frequencies, the dynamical matrix is calculated with the primitive hexagonal cell by using density-functional-perturbation theory\cite{Gajdos2006}
and is processed with the {\small{PHONOPY}} code.\cite{phonopy}
The oscillator strength and dielectric function are calculated by combining Born effective charge tensors and high frequency dielectric constant from the electron response.\cite{Gonze1997,Paudel2007}



Figure \ref{Structure}(e, f) summarizes the infrared response of Co$_4$Nb$_2$O$_9$. A symmetry analysis reveals 7$A_{2u}$ + 14$E_u$  infrared-active modes and 7$A_{1g}$ + 15$E_g$ Raman-active modes.
The doubly-degenerate $E_u$ vibrational modes appear in the $ab$-plane whereas the singly-degenerate $A_{2u}$ modes vibrate along $c$. The Ta analog is isostructural 
with  a $P\bar{3}c1$ space group and  $D_{3d}$ point group symmetry. As a result, the spectrum of Co$_4$Ta$_2$O$_9$ is quite similar to that of the Nb-containing compound [Fig. S1, Supplementary information]. 
Overall, the number of infrared-active modes and their peak positions are nearly identical in both  Co$_4$Nb$_2$O$_9$ and Co$_4$Ta$_2$O$_9$. 
A summary and detailed comparison between the experimental and theoretical phonon frequencies, symmetries, and displacement patterns is available in Table S6, Supplementary information.
The modes related to the heavy Nb and Ta centers appear at the lowest frequencies since $\omega \sim \sqrt{k/\mu}$. We also expect features involving  Nb to vibrate at slightly higher frequencies than those involving Ta due to simple mass effects. 
Here, $\omega$ is the frequency, $k$ is the spring constant, and $\mu$ is the effective mass. 
We can test this supposition by examining the mode displacement patterns and realizing that the $E_u$ modes at 60(58) cm$^{-1}$ and the $A_{2u}$ modes at 138(128) cm$^{-1}$ reflect the presence of Nb vs. Ta, respectively.
The $E_{u}$ symmetry phonon near 150 cm$^{-1}$ that involves shearing of the Co planes against each other will be important in our  discussion as well. This structure is marked with a green arrow in Fig. \ref{Structure}(f). 
We also studied the temperature dependence of the phonons in Co$_4$Nb$_2$O$_9$ and the Ta analog. The majority of features move systematically with decreasing temperature. The $E_u$ symmetry shearing mode near 150 cm$^{-1}$ is the only exception.

\begin{figure*}[tbh]
\begin{minipage}{7in}
\includegraphics[width = 7in]{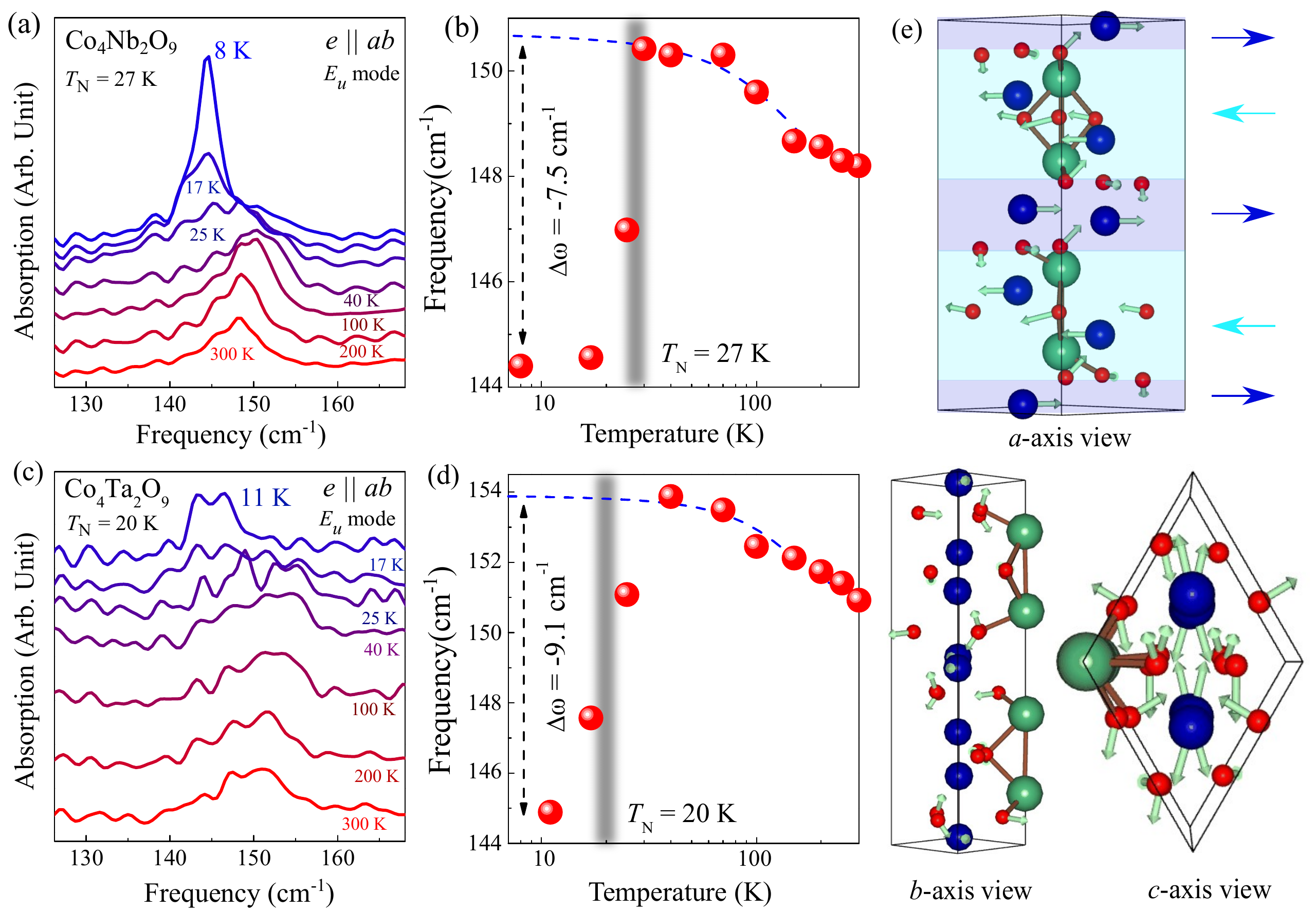}
\end{minipage}
\begin{minipage}{7in}
\caption{\label{SpinPhonon}(a, b) Absorption spectrum of Co$_4$Nb$_2$O$_9$ as a function of temperature and peak position vs. temperature for the 150 cm$^{-1}$ shearing mode. (c, d) Similar results for Co$_4$Ta$_2$O$_9$. The spectra in (a, c) are shifted for clarity, and the blue dashed lines in (b, d) account for temperature effects. 
(e) Calculated displacement pattern of the $E_u$ symmetry Co$^{2+}$ layer shearing mode near 150 cm$^{-1}$. 
}
\end{minipage}
\end{figure*}


Figure \ref{SpinPhonon}(a, c) displays a close-up view of the $E_u$ symmetry vibrational mode near 150 cm$^{-1}$ in both materials. 
This feature hardens with decreasing temperature in the paramagnetic phase and softens across $T_{\rm N}$ in both systems. It is the only mode that displays this behavior.
The Nb compound hosts a sharp frequency shift across the magnetic ordering temperature  whereas the Ta analog reveals a gradual transition with a sluggish frequency shift and noticeable precursor effect. As shown in Fig. \ref{SpinPhonon}(b, d), we fit frequency vs. temperature trends in the paramagnetic phase using a Boltzmann sigmoidal model to capture anharmonic effects:
\begin{equation}\label{Fit}
{\omega}(T) = \frac{(\omega_{o}-\omega_{h})}{(1+e^{((T-T_i)/dT)})}+\omega_{h}.
\end{equation}
Here, $\omega_{h}$ and $\omega_{o}$ are the high and low temperature limits; $T_i$ and $dT$ are the inflection  point and the width of the distribution. 
As demonstrated by the deviation from the anharmonic phonon model fit below the magnetic ordering transition, this phonon engages in strong spin-phonon coupling in  both Co$_4$Nb$_2$O$_9$ and Co$_4$Ta$_2$O$_9$.
\cite{Hughey2017,Oneal2019,Fan2021}
We also employed line width effects to extract phonon lifetimes. Focusing on the Co$^{2+}$ layer shearing mode, we find  phonon lifetimes of 0.75 and 0.6 ps at room temperature in Co$_4$Nb$_2$O$_9$ and the Ta analog, respectively. They drop slightly with decreasing temperature and rise sharply below $T_{\rm N}$. The latter is consistent with fewer scattering events due to spin ordering.\cite{Sun2013,Alessandro2019}  Details are available in the Supplementary information.

In magnetic materials, it is well-known that phonon frequencies can be sensitive to the development of near-neighbor spin correlations.\cite{Granado1999,Sushkov2005,Fennie2006,Birol2013} In such cases, the  frequency is expressed as:\cite{Baltensperger1970} 
\begin{equation}\label{SP}
    \omega = \omega _{o} + \lambda\langle S_i\cdot S_j \rangle.
\end{equation}    
Here, $\omega_{o}$ is the unperturbed phonon frequency that is nicely revealed  from the fit in Fig. \ref{SpinPhonon}(b, d) at base temperature, $\omega$ is the
renormalized phonon frequency due to spin-phonon interactions, $\langle S_i\cdot S_j \rangle$ is the spin-spin correlation function, and $\lambda$ is the spin-phonon coupling constant. 
Using Eqn. \ref{SP}, our frequency vs. temperature trends, and the model fits of ${\omega}$($T$) shown in Fig. \ref{SpinPhonon}(b, d), we extract frequency shifts across the magnetic ordering transitions and calculate spin-phonon coupling constants for Co$_4$Nb$_2$O$_9$ and Co$_4$Ta$_2$O$_9$.

The displacement pattern of the participating phonon is also key to our analysis. As a reminder, our calculations predict  that the 150 cm$^{-1}$ phonon mode is an  out-of-phase $E_u$ symmetry displacement involving the Co centers in which the planar and buckled layers vibrate against each other. 
The Nb and Ta centers do not participate very much in the motion. As a result, the relevant motion contains only one spin-containing center: Co$^{2+}$ with  $S$ = $\frac{3}{2}$. With this insight and the fact that $\langle S_i\cdot S_j \rangle$ goes as $S^2$ in the low-temperature limit, we see that the spin-spin correlation function can be approximated as $\langle S_i\cdot S_j \rangle$ $\sim$ $S^2$ = ($\frac{3}{2}$)$^2$ = $\frac{9}{4}$. Co$_4$Nb$_2$O$_9$ has a frequency shift ($\Delta\omega$) across $T_{\rm N}$ of -7.5 cm$^{-1}$. In other words, the mode softens across the magnetic ordering temperature. This means that the bond between layers gets weaker due to magnetism. We extract a spin-phonon coupling constant  ($\lambda$) of -3.3 cm$^{-1}$. Co$_4$Ta$_2$O$_9$ has a larger frequency shift than Co$_4$Nb$_2$O$_9$. Using the same estimate for $\langle S_i\cdot S_j \rangle$ and $\Delta\omega$ =  -9.1 cm$^{-1}$, we find $\lambda$ = -4.0 cm$^{-1}$. This makes sense because Nb is a 4$d$ element whereas Ta is 5$d$, so the latter hosts  more significant spin-orbit coupling.

We can extend this analysis to include additional effects.
For instance, we can modify the spin-spin correlation function as $\langle S_i \cdot S_j\rangle \approx S^2cos^2(\Theta)$ in order to capture the noncollinearity of the spin states reported in our systems.\cite{Ding2020,Choi2020} This expression is a simple analog of Malus's rule for the polarization of light.\cite{Collett2005} We use $\theta$ = 10.5$^{\circ}$ and 14$^{\circ}$ to obtain $\lambda$ = -3.4 and -4.3 cm$^{-1}$ for Co$_4$Nb$_2$O$_9$ and Co$_4$Ta$_2$O$_9$, respectively.\cite{Ding2020,Choi2020} Accounting for non-collinearity increases spin-phonon coupling constants in the Co$_4B_2$O$_9$ materials by about 10\%. 
These values are an order of magnitude larger than what was extracted for the aforementioned Raman-active modes,\cite{Yadav2022} so we see that coupling with odd- rather than even-symmetry vibrational modes is significantly more important.

\begin{table*}[tbh]
\centering
\footnotesize
\caption{Spin-phonon coupling in Co$_4$Nb$_2$O$_9$ and Co$_4$Ta$_2$O$_9$ compared with representative  oxides. An asterisk (*) indicates estimated values.}
\label{SPComparison}
\begin{ruledtabular}
\begin{tabular}{|c|c|c|c|c|c|c|c|}
{\bf Materials} & {\bf Sites} & {\bf Electronic} & {\bf $\omega_0$} & {\bf $\Delta\omega$} & {\bf $\Delta\omega$/$\omega_0$} & {\bf $\lambda$} & {\bf Refs.} 
\\
{\bf (cm$^{-1}$)} & & {\bf state} & {\bf (cm$^{-1}$)} & {\bf (cm$^{-1}$)} & {\bf (cm$^{-1}$)} & {\bf (cm$^{-1}$)} &   \\
\hline
ZnCr$_2$O$_4$ & Cr$^{3+}$ & 3$d^3$ & 370 & 11 & 3\% & 6.2 & \citenum{Sushkov2005}
\\ 
\hline
CdCr$_2$O$_4$ & Cr$^{3+}$ & 3$d^3$ & 365 & 9 & 2.5\% & 4 & \citenum{ValdesAguilar2008}
\\ 
\hline
SrMnO$_3$ & Mn$^{4+}$ & 3$d^3$ &  165 & 30*
& 18\%*
& 4.8* 
& \citenum{Kamba2014}
\\ 
\hline
Fe$_{1-x}$Cu$_{x}$Cr$_2$S$_4$ & Fe$^{3+}$ & 3$d^5$ & 120 to 400 & - & $<$-1.5 to 3\% & -  & \citenum{Rudolf2005}
\\ 
\hline
Fe$_2$TeO$_6$ & Fe$^{3+}$ & 3$d^{5}$ & 300 to 800 & $<$1 to 5* &  $<$1 to 1.3\%*  & 0.1 to 1.2 & \citenum{Pal2021}\\ 
\hline
Sr$_2$CoO$_4$ & Co$^{4+}$ & 3$d^5$ & 630 and 410 & - & - & 2 to 3.5 & \citenum{Pandey2013}
\\ 
\hline
MnF$_2$ & Mn$^{2+}$ & 3$d^5$ & $\approx$56 to 480.5 & 2, -1.2, 2.7, 1.5* & $\approx$-0.4 to 3.7\%* & 0.4, 0.3, 0.3, -0.2  & \citenum{Lockwood1988}
\\ 
\hline
FeF$_2$ & Fe$^{2+}$ & 3$d^6$ & ~56 to 480.5 & - & - & 0.4, 0.3, -0.5, -1.3  & \citenum{Lockwood1988}
\\ 
\hline
{\bf Co$_4$Nb$_2$O$_9$} & Co$^{2+}$ & 3$d^7$ & 144 & -7.5 & -5\% & -3.4 & {\bf This work} 
\\
\hline
{\bf Co$_4$Ta$_2$O$_9$} & Co$^{2+}$ & 3$d^7$ & 145 & -9.1 & -6\% & -4.3 & {\bf This work} 
\\ 
\hline
NiO & Ni$^{2+}$ & 3$d^{8}$ & 752.5, 1160 & -12.5, 25* &  -1.7\%*, -2.7\%*  & -7.9, 14.7 & \citenum{Aytan2017}\\ 
\hline
Ni$_3$TeO$_6$ & Ni$^{2+}$ & 3$d^{8}$ & 313, 597.3, 672 & -0.4, 0.3, -3.7 &  -0.1 to <1\%*  & -0.4, 0.3, -3.7 & \citenum{Yokosuk2015}\\ 
\hline
Y$_2$Ru$_2$O$_7$ & Ru$^{4+}$ & 4$d^{4}$ & 420 and 492 & -0.1 and -0.3 & -0.2 and -0.6\%* & -6 and -9 & \citenum{Lee2004}\\ 
\hline
NaOsO$_3$ & Os$^{5+}$ & 5$d^3$ & 550 to 800 & 40 & $\approx$5.7\%* & 17.8* & \citenum{Calder2015}
\\ 
\hline
Cd$_2$Os$_2$O$_7$ & Os$^{5+}$ & 5$d^{3}$ & 100 to 800 & -4.0 to 20* & -1 to 7\% & -1.8 to 8.9 (with $S$ = 3/2)* & \citenum{Calder2015,Sohn2017}\\ 
\hline
Y$_2$Ir$_2$O$_7$ & Ir$^{4+}$ & 5$d^{4}$ & 333, 425, 500 & -0.8*, -1.6*, -6.5* & -0.2, -0.4, -1.3\% & -0.4 to -3.2 (with $J_{\rm eff}$ = 1/2)* & \citenum{Son2019}\\ 
\hline
Sr$_3$NiIrO$_6$ & Ir$^{4+}$ & 5$d^4$ & 133, 310, 534 & - & - & 2, 10, 5 & \citenum{Oneal2019}
\\ 
\hline
Ba$_2$FeReO$_6$ & Fe$^{3+}$/Re$^{5+}$ & 3$d^{5}$/5$d^{2}$ & 390 to 630 & $\approx$30 & 5.1\%* & -  & \citenum{GarciaFlores2012}\\ 
\hline
Sr$_2$CrReO$_6$ & Cr$^{3+}$/Re$^{5+}$ & 3$d^{3}$/5$d^{2}$ & 600 & $\approx$25 & 4.9\%* & -  & \citenum{GarciaFlores2012}\\ 
\end{tabular}
\end{ruledtabular}
\end{table*}

We can also analyze the individual interactions between Co$^{2+}$ sites. Here, it's important to recall that Co$_4$Nb$_2$O$_9$ and the Ta analog are composed of two different layers - akin to a superlattice  consisting of planar graphene and buckled SiC [Fig. \ref{Structure}(c, d)].  Writing down the spin Hamiltonian for the planar layer, the buckled layer, and the interaction terms between layers in a pair-wise fashion, we obtain: 
\begin{equation}\label{Hamiltonian}
    {\hatH_{spin} = {\sum_{i,j}J_PS_i{\cdot}S_j + {\sum_{i,j}}J_BS_i{\cdot}S_j + {\sum_{i,j}}J_IS_i{\cdot}S_j}.}
\end{equation}
\noindent
Here, the $J_P$'s are in-plane exchange interactions, the $J_B$'s are those in the buckled layer, and the $J_{I}$'s couple the two layers  quantifying interlayer exchange interactions. The $S_i$'s and $S_j$'s are the spins. 
Interestingly, both ${\partial}^2J_P/{\partial}u_m{\partial}u_n$  and ${\partial}^2J_B/{\partial}u_m{\partial}u_n$ are zero because the motion of interest does not modulate these exchange interactions [Section 4, Supplementary Information]. In other words, the in-plane terms can be ignored because the distances and angles do not change as a result of the displacement. Here, the $u_{n,m}$'s are the displacements (or distances) between Co centers. Writing down the force constant ($k_{n,m}$), we find that: 
\begin{equation}\label{ForceConstant}
    k_{n,m} = {\frac{{{\partial}^2\hatH_{spin}}}{{{\partial}u_m{\partial}u_n}}} = {\frac{{\partial}^2J_{I}}{{\partial}u_m{\partial}u_n}}{\langle}S_i{\cdot}S_j{\rangle} = {\lambda}_{I}{\langle}S_i{\cdot}S_j{\rangle}.
\end{equation}
We therefore see that spin-phonon coupling in Co$_4$Nb$_2$O$_9$ and the Ta analog is entirely an inter-plane effect and that it is the Co shearing mode shown in Fig. \ref{SpinPhonon}(e) that modulates the interlayer magnetic interactions. 
It turns out that there are two primary types of inter-plane interactions in these systems. (We neglect the long interaction between Co centers along $c$ because it is very small.)  By analyzing the bond angles and their tendencies toward parallel or anti-parallel alignment based upon Goodenough-Kanamori-Anderson rules\cite{} along with the number of near-neighbors, we can write: 
%

\begin{equation}\label{CouplingConstant} 
\begin{split}
    \lambda_{Total} \langle S_i \cdot S_j \rangle = \lambda_{I} \langle S_i \cdot S_j \rangle \\
    & \hspace{-0.8in} = \lambda_{2,FM} \langle S_{i} \cdot S_{j} \rangle  +  3\lambda_{3, AFM} \langle S_{i} \cdot S_{j} \rangle.
\end{split}
\end{equation}

%
\noindent
We immediately notice the competition between antiferromagnetic and ferromagnetic interlayer interactions in Eqn. \ref{CouplingConstant} [and Fig. \ref{Structure}(d)], suggesting that Co$_4$Nb$_2$O$_9$ and Co$_4$Ta$_2$O$_9$ are 
frustrated. This competition reduces the overall size of the frequency shift across the magnetic ordering transition. It is also why the $T_{\rm N}$'s are so low. We suspect that the magnitude of ${\lambda}_{2,FM}$ is larger than that of ${\lambda}_{3,AFM}$'s, but there are more ${\lambda}_{3,AFM}$'s in the sum leading to a slight preference for an antiferromagnetic ground state. 
The findings are consistent with softer O--Co--O bond angles and temperature-dependent lattice constants.\cite{Yin2016,Yadav2022}  As we discuss below, magnetoelectric coupling in these materials is likely to emanate from inter-layer spin-phonon interactions.


Table \ref{SPComparison} summarizes the properties of several representative transition metal oxides, 4$d$- and 5$d$-containing systems, and 3$d$-5$d$ hybrids. The entries are grouped by the electronic state. 
We immediately notice that some materials have multiple coupled modes and a tendency toward three-dimensional structure whereas others have only a single spin-phonon coupled mode along with a tendency toward layered or chain-like character. As already discussed, the $E_u$ symmetry Co$^{2+}$ shearing mode near 150 cm$^{-1}$ is the only feature to engage in spin-phonon coupling in the materials of interest here. We also see that the frequency shifts in Table \ref{SPComparison} have both positive and negative signs. Among the materials with coupled modes that soften, Co$_4$Nb$_2$O$_9$ and the Ta analog host the largest relative frequency shift - even though competing antiferromagnetic and ferromagnetic interactions reduce the size of the overall shift. 

The spin-phonon coupling constant, $\lambda$, is often challenging to define, and the procedure for doing so is inconsistent in the literature. We therefore find the relative frequency shifts to be the most reliable for comparison purposes. As pointed out in Ref.~\citenum{Sohn2017}, the relative frequency shift   of a phonon across a magnetic ordering transition $\Delta\omega$/$\omega_0$ is usually less than 1$\%$ for a transition metal oxide. Co$_4$Nb$_2$O$_9$ and the Ta analog are different with relative frequency 
shifts of -5 and -6\%, respectively. 
These values are unusually large. 
In fact, the $\Delta\omega$/$\omega_0$'s that we find in Co$_4$Nb$_2$O$_9$ and Co$_4$Ta$_2$O$_9$ [Fig. \ref{SpinPhonon}(b, d)] are more comparable with the 5$d$ oxides and 3$d$/5$d$ hybrid systems  shown in Table \ref{SPComparison} where the relative frequency shifts correlate (in general) with the electronic shell and  spin-orbit coupling.

The presence of a heavy element at the $B$ site in Co$_4$Nb$_2$O$_9$ and Co$_4$Ta$_2$O$_9$ naturally raises questions about the spin Hamiltonian and whether additional terms such as anisotropy and Dzyaloshinskii-Moriya interactions contribute to spin-phonon coupling.\cite{Lee2004,Silverstein2014,Son2019,Kim2020} Recent neutron scattering also demonstrates that the excitation spectrum can not be reproduced without terms that give rise to spin noncollinearity.\cite{Deng2018,Ding2020} Focusing on Co$_4$Nb$_2$O$_9$, we see that the anisotropies $\geq$ $J$'s $\geq$ Dzyaloshinskii-Moriya interaction.\cite{Deng2018,Ding2020} All of these terms are on the order of 1 meV. 
%
Of course, we are not interested in the absolute size of the anisotropy, exchange, or Dzyaloshinskii-Moriya interaction. Instead, we want to know how they are modulated by the 150 cm$^{-1}$ phonon mode. In other words, we are interested in how these quantities {\it change} with respect to the displacement. By using the mode symmetry and details of the displacement pattern, we can identify the terms that are important for spin-phonon coupling as well as those that will likely cancel out.

If we modify Eqn. \ref{ForceConstant} to include these terms, the prefactors that contribute to spin-phonon coupling are ${\partial}^2A/{\partial}u^2$ and ${\partial}^2DM/{\partial}u^2$. Here $A$ is the anisotropy, and $DM$ is the Dzyaloshinskii-Moriya interaction. The question is whether these contributions are large or small. 
Since $A$ is an on-site term, we do not expect it to change very much with a vibration. 
Even if it does, we anticipate that shearing of the planar vs. buckled 
layers against each other will significantly 
diminish ${\partial}^2A/{\partial}u^2$ due to their opposite motion.  In other words, while ${\partial}^2A/{\partial}u^2$ for the planar and buckled layers are not exactly equal, 
they have opposite signs due to the shearing motion of the layers which, when added together, diminish any impact of the overall anisotropy term. 
Therefore, we argue that in these unique circumstances ${\partial}^2A/{\partial}u^2$ is small.
On the other hand, the ${\partial}^2DM/{\partial}u^2$ term multiplies a cross product between two sites and has the potential to contribute. However, spin-orbit coupling and noncollinearity derive primarily from the $B$ site, suggesting reduced importance because the displacement pattern does not involve the movement of the $B$ site. Another way to consider the issue is that the Dzyaloshinskii-Moriya interaction is in the Co-O-Co linkage of the buckled layers, and these bond lengths and angles do not change with the motion.\cite{Deng2018} This again rules out contributions from the Dzyaloshinskii-Moriya interaction. 

Magnetoelectric coupling in the Co$_4B_2$O$_9$ ($B$ = Nb, Ta) family of materials has been of sustained interest. \cite{Khanh2016,Fang2015,Fang2015,Lee2020,Cao2017}
Suggested mechanisms include spin-phonon coupling,\cite{Xie2016,Khanh2019} 
domain and domain wall effects,\cite{Fang2015} Dzyaloshinskii-Moriya interactions through the spin-current model,\cite{Solovyev2016,Deng2018} and critical spin fluctuations.\cite{Fang2015,Yin2016} 
With the direct observation of large frequency shifts across the magnetic ordering transitions, sizeable spin-phonon coupling constants, and microscopic analysis of the competing interlayer interactions, we are in a better position to evaluate how magnetoelectric coupling might emerge from these candidate mechanisms.
%
%
As established above, our analysis reveals that Co$_4$Nb$_2$O$_9$ and the Ta analog have strong spin-phonon interactions that involve competing interlayer exchange interactions modulated by the shearing motion of the Co layers. While there are Raman-active modes that are sensitive to the development of magnetic ordering,\cite{Yadav2022} the infrared-active $E_u$ symmetry Co$^{2+}$ layer shearing mode has a frequency shift that is an order of magnitude larger, indicating that odd-symmetry motion dominates spin-phonon coupling in this system. This type of exchange striction provides a very natural origin 
for magnetodielectric coupling.\cite{Xie2016,Khanh2016,Solovyev2016,Khanh2019} It is also the most likely origin of magnetoelectric coupling\cite{Xie2016,Khanh2016,Khanh2019} given the fact that the Dzyaloshinskii-Moriya interaction operating through the spin-current model is significantly smaller and does not couple to the  Co$^{2+}$ layer shearing mode near 150 cm$^{-1}$. The change in dipole moment  associated with magnetoelectric coupling\cite{Spaldin2021} 
is consistent with the microscopic nature of this infrared-active mode.

To summarize, we measured the vibrational properties of Co$_4$Nb$_2$O$_9$ and the Ta analog and compared our findings with lattice dynamics calculations and a detailed model of spin-phonon 
coupling. 
In addition to revealing one of the largest relative frequency shifts ever reported, these materials host a Co shearing mode that couples only with the interlayer interactions due to unique symmetry conditions. 
These interlayer interactions are frustrated. 
%
%
Given the sizable contribution of spin-phonon interactions in these systems, it is likely that magnetoelectric coupling is driven by this effect.

\section*{Supplementary Information}

See Supplementary information for the complete description of the vibrational modes, displacement patterns, the temperature dependence of phonons, spin-phonon analysis, and phonon lifetimes as a function of temperature.

\subsection*{}
\vspace{-0.5in}
Research at the University of Tennessee is supported by the U.S.
Department of Energy, Office of Basic Energy Sciences, Materials Science Division under award DE-FG02-01ER45885. SWC was supported by the center for Quantum Materials Synthesis (cQMS), funded by the Gordon and Betty Moore Foundation’s EPiQS initiative through GBMF10104 and by Rutgers University.
D.V. was supported by the NSF (DMR-1954856).
V.K. was supported by the National Science Foundation (DMR-2103625).
S.C. was supported by the Institute for Basic Science (IBS-R011-Y3).
D. G. Oh, N. Lee, and Y. J. Choi were supported by the National Research Foundation of Korea (NRF-2017R1A5A1014862 (SRC program: vdWMRC center), NRF-2019R1A2C2002601, and NRF-2021R1A2C1006375).
J.H.L. at UNIST was supported by Midcareer Researcher
(2020R1A2C2103126) and Creative Materials Discovery
(2017M3D1A1040828) programs through the National
Research Foundation of Korea.

\section{Author Declarations}
\subsection*{Conflict of Interest}
The authors have no conflicts to disclose.

\subsection*{Author Contributions}

KP and JLM designed the study. CK, DGO, NL, SWC, and YJC grew the crystals. KP performed the variable temperature measurements. KP and JLM analyzed the spectral data. JK and DV performed DFT calculations. KP, JK, SC, SF, VK, and JLM discussed the measurement results and analysis in detail. KP, JHL, and JLM developed a spin Hamiltonian model. KP, JK, and JLM wrote the manuscript. All authors read and commented on the text.

\section*{Data Availability}

The data that support the findings of this study are available from the corresponding author upon reasonable request.

\section*{References}

\bibliographystyle{apsrev4-1}
\bibliography{citation}

\end{document}